\documentclass[preprintnumbers,amsmath,amssymb,11pt]{revtex4}
\usepackage{graphicx}
\usepackage{dcolumn}
\usepackage{bm}
\begin{document}
\title{Intermediate Inflation in the Jordan-Brans-Dicke Theory}
\author{Antonella Cid}
\email{acidm@ubiobio.cl	}
\affiliation{Departamento de F\'{\i}sica, Universidad del B\'io-B\'io, Casilla 5-C, Concepci\'{o}n, Chile}
\author{Sergio del Campo}
\email{sdelcamp@ucv.cl	}
\affiliation{Instituto de F\'{\i}sica, Pontificia Universidad Cat\'olica de Valpara\'iso, Valpara\'iso, Chile}

\begin{abstract}
We present an intermediate inflationary stage in a Jordan-Brans-Dicke theory. In this scenario we analyze the quantum fluctuations corresponding to adiabatic and isocurvature modes. The model is compared to that described by using the intermediate model in Einstein General Relativity theory. We assess the status of this model in light of the WMAP7 data.
\end{abstract}
\maketitle
\section{Introduction}

The inflationary paradigm \cite{inflation,Linde} has been confirmed as the most successful candidate for explaining the physics of the very early universe \cite{KKMR}. This sort of scenarios solves some of the puzzles of the standard cosmological model, such as the horizon, flatness and entropy problems, as well as providing for a mechanism to seed structures in the universe.

We study a particular scenario called intermediate inflation \cite{Barrow1}, characterized for the scale factor evolving as $a(t)=a_0\exp(A t^f)$. In this model, the expansion of the universe is slower than standard de Sitter inflation ($a(t)=\exp(H t)$), but faster than power law inflation ($a(t)= t^p,\ p>1$). The intermediate inflationary model was first introduced as an exact solution corresponding to a particular scalar field potential of the type $V(\phi)\propto \phi^{-4(f^{-1}-1)}$ in the slow-roll approximation, where $0<f<1$.

On the other hand, there has been carried out a less standard theory of gravity, namely scalar-tensor theory of gravity \cite{JBD,BEPA}. The archetypical theory associated with scalar-tensor models is Jordan-Brans-Dicke \cite{JBD}. The JBD theory is a class of model in which the effective gravitational coupling evolves with time. The strength of this coupling is determined by a scalar field, the so-called JBD field, which tends to the value $G^{-1}$. In modern context, JBD theory appears naturally in supergravity models, Kaluza-Klein theories and in all known effective string actions \cite{bla}. 

We present an intermediate inflationary universe model in a JBD theory. We will write the Friedmann field equations, together with the corresponding scalar field equations. The intermediate inflationary period of inflation will be consistently described in the slow-roll approximation. Scalar and tensor perturbations will be expressed in terms of the parameters that appear in our model and these parameters will be constrained by taking into account the WMAP seven year data \cite{WMAP2}. 

\section{Background Equations in the Einstein Frame}

A wide class of non-Einstein gravity models can be recast in the action \cite{Starobinsky2}:
\begin{eqnarray}
\label{BDI1}
S=\int \sqrt{-g}d^4x\left[\frac{1}{2\kappa^2}R+\frac{1}{2}g^{\mu\nu}\partial_{\mu}\chi\partial_{\nu}\chi+\frac{1}{2}e^{-\gamma \kappa \chi}g^{\mu\nu}\partial_{\mu}\phi\partial_{\nu}\phi-e^{-\beta\kappa\chi}V(\phi)\right],
\end{eqnarray}
where $R$ is the Ricci scalar, $\kappa^2=8\pi G$, $\beta$ and $\gamma$ are constants, $\chi$ and $\phi$ are the dilaton and inflaton fields, respectively. We have considered $c=1$ and the JBD theory is recovered for $\beta=2\gamma$.

We consider a flat Friedmann-Robertson-Walker (FRW) metric in order to obtain the field equations from the action (\ref{BDI1}). These equations in the slow-roll regime are given by \cite{Anto}:
\begin{eqnarray}
\label{BDI13}
3H\dot{\chi}-\beta\kappa e^{-\beta\kappa\chi} V(\phi)=0,\ \ \ \ \ \ 
3H\dot{\phi}+e^{(\gamma-\beta)\kappa\chi}V'(\phi)=0,\ \ \ \ \ \ \label{BDI15}
3H^2-\kappa^2e^{-\beta\kappa\chi}V(\phi)=0.
\end{eqnarray}

By choosing the ansatz $V(\phi)=V_0\phi^n$ in the Jordan-Brans-Dicke theory we can find a solution to the set of Eqs.(\ref{BDI13}):
\begin{eqnarray}
\label{BDI16}
\chi=\frac{\beta}{\kappa}\ln\left(\frac{a}{a_b}\right)+\chi_b,\ \ \ \ \ a(t)=a_b\left(\frac{1+\frac{A}{p} t^f}{1+\frac{A}{p} t_b^f}\right)^{p}
\ \ \ \ \ \textrm{and}\ \ \ \ \ 
\phi(t)=\left(\frac{8\sqrt{V_0}(1-f) t}{\sqrt{3}\kappa f^2}\right)^{\frac{f}{2}}
\end{eqnarray}
where the subscript $b$ denotes values at the beginning of the inflationary epoch, $p\equiv\frac{2}{\beta^2}$, $f\equiv\frac{4}{4-n}$ and $A$ is an integration constant. 

We note that the scale factor in Eq.(\ref{BDI16}) is a generalization of the scale factor corresponding to intermediate inflation in the Einstein theory \cite{Barrow1}, in the case $p\rightarrow\infty$ we recover $a(t)\propto e^{At^f}$. The authors of Ref.\cite{LaSteinhardt} found the same form for $a(t)$ when they first studied a cosmological model in a JBD theory, named extended inflation.

Observational measurements \cite{ObsWill,ObsBertotti} constraint the parameters $\gamma$ and $\beta$ to be very small in the JBD theory. Furthermore, the field $\chi$ remains very close to a constant after inflation, in the radiation and matter domination eras \cite{JBD}. In order to recover the value of the newtonian gravitational constant after inflation we will consider $\chi$ equals to zero at the end of inflation.

It is well known  that an intermediate stage of inflation needs an additional mechanism to bring inflation to an end \cite{Barrow3}. We will consider that this mechanism starts after $N_T$ e-folds since the beginning of inflation. We normalize $\chi$ in such a way that after $N_T$ e-folds the value of the  field $\chi$ becomes zero, therefore $\chi_b=-\frac{\beta}{\kappa}N_T$. We assume that the value of $\chi$ remains zero after that time.

\section{Linear Order Scalar Perturbations}

We analyze the cosmological scalar perturbations in the longitudinal gauge \cite{Mukhanov}. The perturbed Einstein field equations in the slow-roll regime and for non-decreasing adiabatic and isocurvature modes on large scales $k\ll aH$ are given by \cite{Anto}:
\begin{eqnarray}
&&\Phi=\frac{\kappa^2}{2H}\left(\dot{\chi}\delta\chi+e^{-\gamma\kappa\chi}\dot{\phi}\delta\phi\right)=\frac{\beta\kappa}{2}\delta\chi-\frac{V'(\phi)}{2V(\phi)}\delta\phi,\label{BDI25}\\
&&3H\dot{\delta\chi}+(\beta\kappa)^2e^{-\beta\kappa\chi}V(\phi)\delta\chi-\beta\kappa e^{-\beta\kappa\chi}V'(\phi)\delta\phi=2\beta\kappa e^{-\beta\kappa\chi}V(\phi)\Phi,\label{BDI26}\\
&&3H\dot{\delta\phi}+e^{(\gamma-\beta)\kappa\chi}V''(\phi)\delta\phi+(\gamma-\beta)\kappa V'(\phi)e^{(\gamma-\beta)\kappa\chi}\delta\chi=-2e^{(\gamma-\beta)\kappa\chi}V'(\phi)\Phi.\label{BDI27}
\end{eqnarray}

The authors in Ref.\cite{Starobinsky,Chiba-Sugiyama-Yokoyama} have found the solution to the set of Eqs.(\ref{BDI25})-(\ref{BDI27}) to be:
\begin{eqnarray}
&&\frac{\delta\chi}{\dot{\chi}}=\frac{C_1}{H}-\frac{C_3}{H},\label{BDI28}\ \ \ \ \ \ 
\frac{\delta\phi}{\dot{\phi}}=\frac{C_1}{H}+\frac{C_3}{H}(e^{-\gamma\kappa\chi}-1),\\
&&\Phi=-C_1\frac{\dot{H}}{H^2}+C_3\left(\frac{(1-e^{\gamma\kappa\chi})}{2\kappa^2}\left(\frac{V'(\phi)}{V(\phi)}\right)^2-\frac{\beta^2}{2}\right),\label{BDI30}
\end{eqnarray}
where $C_1$ and $C_3$ are two integration constants related with the initial values of $\delta\phi$, $\delta\chi$, $\phi$ and $\chi$. The terms proportional to $C_1$ and $C_3$ represent adiabatic and isocurvature modes, respectively\cite{Starobinsky2}. The isocurvature nature of the term proportional to $C_3$ is guaranteed by the fact that the second term in Eq.(\ref{BDI30}) is vanishingly small after inflation when $\chi(t)\approx0$. 

In order to get the spectrum of scalar perturbations \cite{Garcia-Bellido-Wands} we  calculate the comoving curvature perturbation, ${\cal R}$, which in this case turns to be \cite{Anto}:
\begin{eqnarray}
\label{BDI32}
{\cal R}=C_1-C_3W,\ \ \ \ \textrm{with} \ \ \ W=1-\left(e^{\gamma\kappa\chi(t)}+\beta^2\kappa^2\left(\frac{V(\phi)}{V'(\phi)}\right)^2\right)^{-1}
\end{eqnarray}

As we see from Eq.(\ref{BDI32}), the term $W$ is responsible for the change of ${\cal R}$ during the inflationary stage. For intermediate inflation in a JBD theory we can calculate $W$ for the chosen potential \cite{Anto}:
\begin{eqnarray}
\label{BDI33}
W(N)=1+\frac{e^{\frac{N_T\beta^2}{2}}(1-f)(2-\beta^2)}{2f-\beta^2-e^{\frac{N\beta^2}{2}}(2-\beta^2)}.
\end{eqnarray}
Here we have used the number of e-folds $N$ to describe time evolution because it is more convenient in the subsequent analysis. We note that the case $\beta=0$ corresponds to $W=0$, which is expected because for $\beta=0$ there is only one scalar field driving inflation, and consequently the comoving curvature perturbation $\cal{R}$ remains constant on large scales \cite{Mukhanov}.

The current observational constraints bring $\beta$ to an upper limit given by $\beta \le0.02$ \cite{ObsBertotti}. There exist a wide range of $N_T$ for which we can find values for $\beta$ and $f$ in the allowed ranges in such a way that $\vert W\vert<0.1$. For example, for $\beta=0.01$, $f=0.8$ and $N_T\le250$ we get $\vert W\vert<0.05$.

Given that the constants $C_1$ and $C_3$ are related to the initial values of the perturbed fields it is expected they have to be of the same order \cite{Starobinsky2}, then to impose $\vert W\vert<0.1$  guarantees that the variation of ${\cal R}$ during inflation due to the presence of  isocurvature perturbations is small, i.e. for a given $\beta$ and $f$ we can find a maximum $N_T$ value for which $\vert W\vert<$ desired value.

In the following we will consider that ${\cal R}$ remains constant after a given scale $k$ leaves the Hubble horizon during inflation. Furthermore, we will assume that the mechanism to finish inflation does not modify this result.

In Ref.\cite{Anto} it is showed that the error in consider the slow-roll approximation instead of the exact solution in calculating $\mathcal{R}$ is very small.
\section{Spectrum of Curvature and Tensor Perturbations}

The spectrum of the comoving curvature perturbation ${\cal R}$  is given by \cite{Anto}:
\begin{eqnarray}
\label{BDI37}
P_{{\cal R}}(k_*)=\frac{4\pi k_*^3}{(2\pi)^3}\langle \vert C_1 \vert^2\rangle=\left[\frac{H^2e^{2\gamma\kappa\chi}}{(2\pi)^2}\left(\left(e^{-\gamma\kappa\chi}-1\right)^2\frac{H^2}{\dot{\chi}^2}+\frac{H^2e^{\gamma\kappa\chi}}{\dot{\phi}^2}\right)\right]_{t_*}.
\end{eqnarray}
where the expectation values of the scalar field perturbations $\delta\phi$  and $\delta\chi$ are given by random gaussian variables when they cross outside the Hubble radius ($k\approx a_*H_*$) \cite{Mukhanov}.

We note that the presence of a second scalar field during inflation modifies the form of the standard spectrum \cite{Starobinsky2}, the standard form is recovered when we take the limit $\gamma\rightarrow0$. 

The scale dependence of the spectrum is characterized by the spectral index $n_s(k)$ whereas the scale dependence of the spectral index is given by the running $\alpha_s$ \cite{Liddle-Lyth-0}, in our model we have:
\begin{eqnarray}
\label{BDMix}
n_s(k_*)&\equiv&1+\frac{d\ln P_{{\cal R}}}{d\ln k}=1-4\epsilon+\eta+\frac{\beta ^2}{2} Z_1,\\
\alpha_s(k_*)&\equiv&\frac{d n_s}{d\ln k}=-8\epsilon^2+5\epsilon\eta-\xi^2+\frac{\beta ^4}{4} Z_2,
\label{BDIMM7}
\end{eqnarray}
where $\epsilon$ and $\eta$ are the standard slow-roll parameters and $\xi^2$, $Z_1$, $Z_2$ are functions of these parameters defined in Ref.\cite{Anto}. Eqs.(\ref{BDMix})-(\ref{BDIMM7}) reduce to the result in the Einstein theory when $\beta\rightarrow0$ \cite{Barrow3}.

In addition to the scalar curvature perturbations, tensor perturbations can also be generated from quantum fluctuations during inflation \cite{Mukhanov}. The tensor perturbations do not couple to matter and consequently they are only determined by the dynamics of the background metric, so the standard results for the evolution of tensor perturbations of the metric remains valid. The two independent polarizations evolve like minimally coupled massless fields with spectrum \cite{Mukhanov}: 
\begin{eqnarray}
\label{BDIMM8}
P_{{\cal T}}(k_*)=8\kappa^2\left(\frac{H}{2\pi}\right)^2_{t_*}.
\end{eqnarray}

From Eq.(\ref{BDI37}) and Eq.(\ref{BDIMM8}) we can determine the tensor to scalar ratio $r$:
\begin{eqnarray}
r(k_*)=\frac{P_{{\cal T}}}{P_{{\cal R}}}=8\kappa^2\left
[e^{2\gamma\kappa\chi}\left(\left(e^{-\gamma\kappa\chi}-1
\right)^2\frac{H^2}{\dot{\chi}^2}+\frac{H^2e^{\gamma\kappa\chi}}
{\dot{\phi}^2}\right)\right]_{t_*}^{-1},
\end{eqnarray}

\begin{figure}[ht!]
\centering
\includegraphics[scale=0.7]{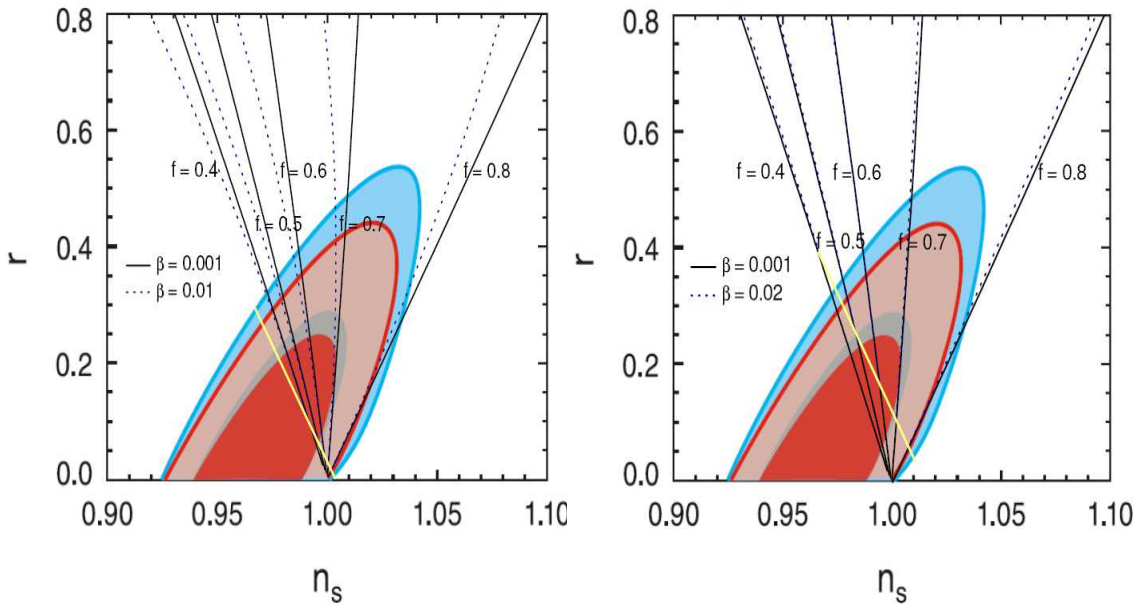}
\caption{Trajectories for different values of the parameter $\beta$ and $f$ in the $n_s-r$ plane. We compare with the WMAP data (five and seven years). The two contours correspond to the $68\%$ and $95\%$ CL \cite{WMAP2}. The left panel shows $N_T=250$ e-folds for $\beta=0.01$ whereas the right panel is for $N_T=60$ e-folds for $\beta=0.02$.}
\label{fig2}
\end{figure}

FIG. \ref{fig2} shows the dependence of the tensor  to scalar ratio on the spectral index for different values of the parameters $\beta$, $f$ and the corresponding maximum $N_T$. For $\beta\le0.001$ there is not significant difference with the case of intermediate inflation in the Einstein theory \cite{Barrow3}.

For $\beta=0.01$, $0.4\le f<0.8$ is well supported by  the data. But there exist a theoretical limit for the model in the maximum number of e-folds of inflation allowed (represented by the yellow line in the left panel of FIG. \ref{fig2}), $N_T\le250$.

In the case $\beta=0.02$, the maximum value for $\beta$ constrained from tests of general relativity \cite{ObsWill,ObsBertotti}, we can have at most 60 e-folds of inflation  in order to have a comoving curvature perturbation close to a constant. This constraint in the number of e-folds exclude $f=0.4$ to be supported by the data  but it still allow $0.5\le f<0.8$ to be well supported.

We see from FIG. \ref{fig2} that the curve $r=r(n_s)$  enters the $95\%$ confidence region for $r\le0.41$ which in terms of the number of e-folds (at the time when a given scale leaves the horizon) means $N>78$ for $f=0.4$, $N>48$ for $f=0.5$, $N>29$ for $f=0.6$ and $N>16$ for $f=0.7$. There are not significant differences for the values of $\beta$ considered in FIG. \ref{fig2}. On the other hand, we have to consider at least 50 e-folds of inflation to push the perturbations to observable scales \cite{Liddle-Lyth-0}, which seems to exclude models for $\beta=0.02$ and $f<0.7$.

\section{Conclusion}

We have studied in detail the intermediate inflationary scenario in the context of a JBD theory. This study was realized in the Einstein frame, but the physical results have to be interpreted in the Jordan physical frame. In this respect, it has been considered that both frames are equivalent, providing that the JBD field varies extremely slowly in the post-inflationary stage of the universe \cite{Starobinsky2}. In this way, the adiabatic fluctuations and the tensor perturbations are described equally in both frames. This allows to obtain explicit expressions for the corresponding power  spectrum of the curvature perturbations $P_{{\cal R}}$, tensor perturbation $P_{{\cal T}}$, tensor-scalar ratio $r$, scalar spectral index $n_s$, and its running $\alpha_s$.

In this work the aim has been to study which set of parameters $\beta$, $f$ and $N_T$ allow us to get a dominant contribution of the adiabatic mode to the power spectrum of scalar perturbations. In order to do that we have restricted the maximum number of e-folds allowed by the model, $N_T$, for a given set of parameters $\beta$ and $f$. For a given value of $\beta$ and $0<f<1$ we get ${\cal R}\approx C_1$ constant for a specific value of $N_T$ provided the desired precision. On the other hand, we have restricted ourselves to $\beta\le0.02$ in light of the previous observational constraints on $\beta$ \cite{ObsWill,ObsBertotti}.

We had checked numerically that the slow-roll approximation is adequated, even to the analysis of first order perturbations, this is valid for the range of parameters considered in this work \cite{Anto}.

In order to bring some explicit results we have taken the constraint in the $n_s-r$ plane coming from the seven-year WMAP data. We have found that the parameter $f$, which initially lies in the range $0<f<1$ for this model, is well supported by the data as could be seen from FIG.\ref{fig2}.

On the other hand we have to consider at least 50 e-folds of inflation to push the perturbations to observable scales \cite{Liddle-Lyth-0}, which seems to exclude models for $\beta=0.02$ and $f<0.7$. Thus, we see that our study has allowed us to put restrictions on the parameters that appear in our model by comparing to the WMAP7 results in terms of $n_s-r$ plane. We have not considered in this work the incidence of the running of the spectral index in the constraints of the model.

Finally in this work, we have not addressed the phenomena of reheating and possible transition to the standard cosmological scenario. A possible calculation for the reheating temperature would give new constraints on the parameters of our model. We hope to return to this point in the future.

\begin{acknowledgments}
A.C. thanks the Physics Department of Universidad del Bio-Bio for their full support to attend to the first CosmoSul: Cosmology and Gravitation in the Southern Cone.
\end{acknowledgments}


\begin{thebibliography}{99}

\bibitem{inflation}A. Guth, Phys. Rev. D {\bf 23}  347 (1981); A. Albrecht and P.J. Steinhardt, Phys. Rev. Lett. {\bf 48} 1220 (1982).

\bibitem{Linde} A complete description of inflationary scenarios can be found in the book by A. Linde, Particle  Physics and Inflationary
Cosmology, Harwood (1990), arXiv:0503203 [hep-th].

\bibitem{KKMR} W. H. Kinney, E. W. Kolb, A. Melchiorri, and A. Riotto, Phys. Rev. D {\bf  78} 087302 (2008).

\bibitem{Barrow1} J. D Barrow, Phys. Lett. B {\bf 235} 40 (1990); 
									J. D Barrow and P. Saich, Phys. Lett. B {\bf 249} 406 (1990);
									A. Muslimov, Class. Quantum Grav. {\bf 7} 231 (1990); 
									A. D. Rendall, Class. Quantum Grav. {\bf 22} 1655 (2005).

\bibitem{JBD}P. Jordan, Z. Phys. {\bf 157}  112 (1959); C. Brans and R.H. Dicke, Phys. Rev. {\bf 124}  925 (1961).

\bibitem{BEPA} B. Boisseau, G. Esposito-Farese, D. Polarski and A.A. Starobinsky, Phys. Rev. Lett. {\bf 85}  2236 (2000).

\bibitem{bla}P. G. O. Freund, Nucl. Phys. B {\bf 209} 146 (1982); T. Appelquist, A. Chodos and P.G.O. Freund, Modern Kaluza-Klein theories, Addison-Wesley (1987); E.S. Fradkin and A. A. Tseytlin, Phys. Lett. B {\bf  158}  316 (1985); E. S. Fradkin and A. A. Tseytlin, Nucl. Phys. B {\bf  261}  1 (1985); C.G. Callan Jr., E.J. Martinec, M.J. Perry and D. Friedan, Nucl. Phys. B {\bf  262}  593 (1985); C.G. Callan Jr., I.R. Klebanov and M.J. Perry, Nucl. Phys. B {\bf  278}  78 (1986); M.B. Green, J.H. Schwarz and E. Witten, Superstring theory, Cambridge Monographs On Mathematical Physics, Cambridge University Press (1987).

\bibitem{WMAP2} D. Larson et al., Astrophys. J. Suppl. {\bf192} 14 (2011).

\bibitem{Starobinsky2} A.A. Starobinsky, J. Yokoyama, arXiv:9502002 [gr-qc].

\bibitem{Anto} A. Cid, S. del Campo, JCAP 1101 (2011) 013.

\bibitem{LaSteinhardt} D. La  and P. J. Steinhardt, Phys. Rev. Lett. {\bf 62}, 376 (1989).

\bibitem{ObsWill} C. Will : The Confrontation between General Relativity  and Experiment, Living Reviews in Relativity 2001, arXiv:0103036 [gr-qc/].

\bibitem{ObsBertotti} B. Bertotti, L. Iess and P. Tortora, Nature {\bf 425} 374 (2003).

\bibitem{Barrow3} J. D. Barrow, A. R. Liddle and C. Pahud, Phys. Rev. D {\bf 74} 127305 (2006).

\bibitem{Mukhanov} V. F. Mukhanov, H. A. Feldman and R. H. Brandenberger, Phys. Rep. {\bf 215} 203 (1992).

\bibitem{Starobinsky} A. A. Starobinsky , S. Tsujikawa and J. Yokoyama, Nucl. Phys. B {\bf 610} 383 (2001).

\bibitem{Chiba-Sugiyama-Yokoyama} T. Chiba , N. Sugiyama and J. Yokoyama, Nucl. Phys. B {\bf530} 304 (1998).

\bibitem{Garcia-Bellido-Wands} J. Garcia-Bellido and D. Wands, Phys. Rev. D {\bf 53} 5437 (1996).

\bibitem{Liddle-Lyth-0} D. H. Lyth and A. R. Liddle, The Primordial Density Perturbation, Cambridge University Press (2009).

\end{thebibliography}
\end{document}